\documentclass[usenatbib,useAMS]{mnras}

\usepackage{amsmath}
\usepackage{bm}

\usepackage[T1]{fontenc}
\usepackage{ae,aecompl}

\usepackage{multicol}
\usepackage{graphicx}
\usepackage{pdflscape}
\usepackage{color}
\usepackage{xcolor}

\newcommand{\codefont}[1]{{\texttt{#1}}}
\newcommand{\Mpch}{\ensuremath{h^{-1}\mathrm{\,Mpc}}}

\title[Field Level Inference of Massive Structures]{Towards Accurate Field-Level Inference of Massive Cosmic Structures}

\author[S. Stopyra et al.]{Stephen Stopyra$^{1}$\thanks{Contact e-mail: \href{mailto:stephen.stopyra@fysik.su.se}{stephen.stopyra@fysik.su.se}},
	Hiranya V. Peiris$^{1,2}$, Andrew Pontzen$^{2}$, Jens Jasche$^{1}$, Guilhem Lavaux$^{3}$
    \\
    $^{1}$The Oskar Klein Centre for Cosmoparticle Physics, Department of Physics,\\ Stockholm University, AlbaNova, Stockholm SE-106 91, Sweden\\
	$^{2}$Department of Physics and Astronomy, University College London, Gower Street, London WC1E 6BT, UK\\
    $^{3}$ Sorbonne Université, CNRS, UMR 7095, Institut d’Astrophysique de Paris, 75014 Paris, France
    }

\date{Accepted XXX. Received YYY; in original form ZZZ}
\pubyear{2022}

\begin{document}
\label{firstpage}
\pagerange{\pageref{firstpage}--\pageref{lastpage}}
 \maketitle

	\begin{abstract}
		We investigate the accuracy requirements for field-level inference of cluster and void masses using data from galaxy surveys. We introduce a two-step framework that takes advantage of the fact that cluster masses are determined by flows on larger scales than the clusters themselves. First, we determine the integration accuracy required to perform field-level inference of cosmic initial conditions on these large scales, by fitting to late-time galaxy counts using the Bayesian Origin Reconstruction from Galaxies (\codefont{BORG}) algorithm. A 20-step \codefont{COLA} integrator is able to accurately describe the density field surrounding the most massive clusters in the Local Super-Volume ($<135\,\Mpch$), but does not by itself lead to converged virial mass estimates. Therefore we carry out `posterior resimulations', using full $N$-body dynamics while sampling from the inferred initial conditions, and thereby obtain estimates of  masses for nearby massive clusters. We show that these are in broad agreement with existing estimates, and find that mass functions in the Local Super-Volume are compatible with $\Lambda$CDM.

	\end{abstract}


	\section{Introduction}
	\label{sec:introduction}
	
	Traditionally, cosmological constraints have relied on observables constructed from summary statistics of the density field such as the power spectrum. By contrast, the technique of \emph{field-level inference} --- in which the full posterior distribution of the density field is sampled --- potentially allows one to access additional information contained e.g. in the phases of the density field. Examples of the application of field-level inference include a determination of the local matter density from the 2M++ galaxy catalogue with the Bayesian Origin Reconstruction from Galaxies (\codefont{BORG}) algorithm~\citep{jasche2019physical}, the inference of the COSMOS initial density field using Lyman--$\alpha$ data by~\citet{horowitz2019tardis,2019A&A...630A.151P} and~\citet{ata2022predicted}, and the use of effective field theory by~\citet{babic2022bao} to infer the density field on the baryon acoustic oscillation scale from halo catalogues. Field-level inference has also been demonstrated to outperform two-point statistics for weak lensing~\citep{2022MNRAS.509.3194P,porqueres2023field}, and its robustness has previously been investigated in the context of effective-field-theory (EFT) likelihoods~\citep{2021JCAP...03..058N,2022arXiv221207875K}.
	However, accurate field-level inference on scales which are even mildly non-linear at late times is challenging for a number of reasons. The dynamic range of gravitational collapse and the astrophysical complexity of galaxy biasing are key issues that must be addressed in order to accurately infer a density field from observational data. 
	
	As an example of the potential power of field-level inference, cluster masses have long been envisioned as a probe of the cosmological parameters \citep{bocquet2016halo,costanzi2019methods,pratt2019galaxy} and of beyond-$\Lambda$-Cold-Dark-Matter ($\Lambda$CDM) physics such as primordial non-Gaussianity ~\citep{loverde2011non,sartoris2010potential,stopyra2021quantifying} or modified gravity~\citep{mak2012constraints,ilic2019cluster}. However, clusters are structures on scales of order $\mathrm{Mpc}$ which is very small compared with the overall volume in which the inference takes place. Therefore, directly inferring cluster masses using field-level inference within a traditional Bayesian sampling framework would require spatial (and timestepping) resolution that remains, for now, computationally intractable. 
	
	The mass of clusters is nonetheless expected to be physically dictated by large-scale flows~\citep{1985ApJS...58...39B,10.1093/mnras/stz2599,10.1093/mnras/stac1833}. Density and velocity information at such scales can be accurately inferred with currently available approximate dynamical models such as \codefont{FastPM}~\citep{feng2016fastpm} or CO-moving Lagrangian Acceleration~\citep[\codefont{COLA}]{tassev2013solving}. This opens up the possibility of using field-level inference to accurately infer the relevant initial conditions on larger scales. One may then resimulate with high time and spatial resolution a number of samples from the posterior on the initial density field. In this way, a posterior on the mass of a cluster as implied by the combination of larger scale information and the gravity solver can be determined. We refer to this technique as \emph{posterior resimulation} because it takes samples from the posterior distribution of initial conditions, and evolves each to redshift $z=0$ with a higher-accuracy gravity solver.
	
	Posterior resimulation is similar in spirit to the local universe simulations of~\citet{2013MNRAS.435.2065H}. These belong in the broader landscape of local universe simulations, which has its genesis in ~\citet{1989ApJ...344L..53P}, who first studied initial conditions for simulations which closely resemble the local Universe. Local universe simulations are performed with a variety of techniques. For instance, the CLUES collaboration~\citep{gottlober2010constrained,2014MNRAS.437.3586S,2016MNRAS.455.2078S} use Hoffman-Ribak~\citep{hoffman1991constrained} and the Reverse Zel'dovich Approximation~\citep{2013MNRAS.430..888D} to incorporate a number of velocity constraints. A Hoffman-Ribak approach is also used by~\citet{2010MNRAS.406.1007L}. The SIBELIUS simulations~\citep{mcalpine2022sibelius,sawala2022sibelius} use field-level inference to set a large-scale environment, and subsequently introduce additional small-scale power which is necessary to reproduce local group structures. A similar approach is taken with the HESTIA hydrodynamical simulation suite~\citep{2020MNRAS.498.2968L}, in which unconstrained small-scale modes are randomly seeded, and an ensemble of simulations with regions resembling the Local Group is obtained by selecting initial conditions which satisfy a set of criteria on the positions of cluster such as Virgo, and the locations/masses of Local Group galaxies. 	
	
	In contrast to most local universe simulations, \emph{posterior resimulation} uses initial conditions drawn directly from the posterior distribution, and therefore accurately projects statistical uncertainties into the evolved universe, allowing us to explicitly assess the significance of inferred structures. This approach therefore offers a new probe of cluster masses and of other cosmological structure formation observables that are determined by information that resides at larger scales in the initial conditions, and is therefore strongly constrained by the combination of the gravitational collapse process and the large-scale environment. As such, posterior resimulation opens up new avenues for cosmological tests, e.g. one may compare the inferred cluster masses with independent estimates --- from Sunyaev-Zel'dovich (SZ), X-rays or lensing for example --- and for modelling or testing models for galaxy intrinsic alignments, which are believed to be sensitive to large-scale tidal fields~\citep{codis2015spin}.

    \begin{figure*}
		\centering
		\includegraphics[width=\textwidth]{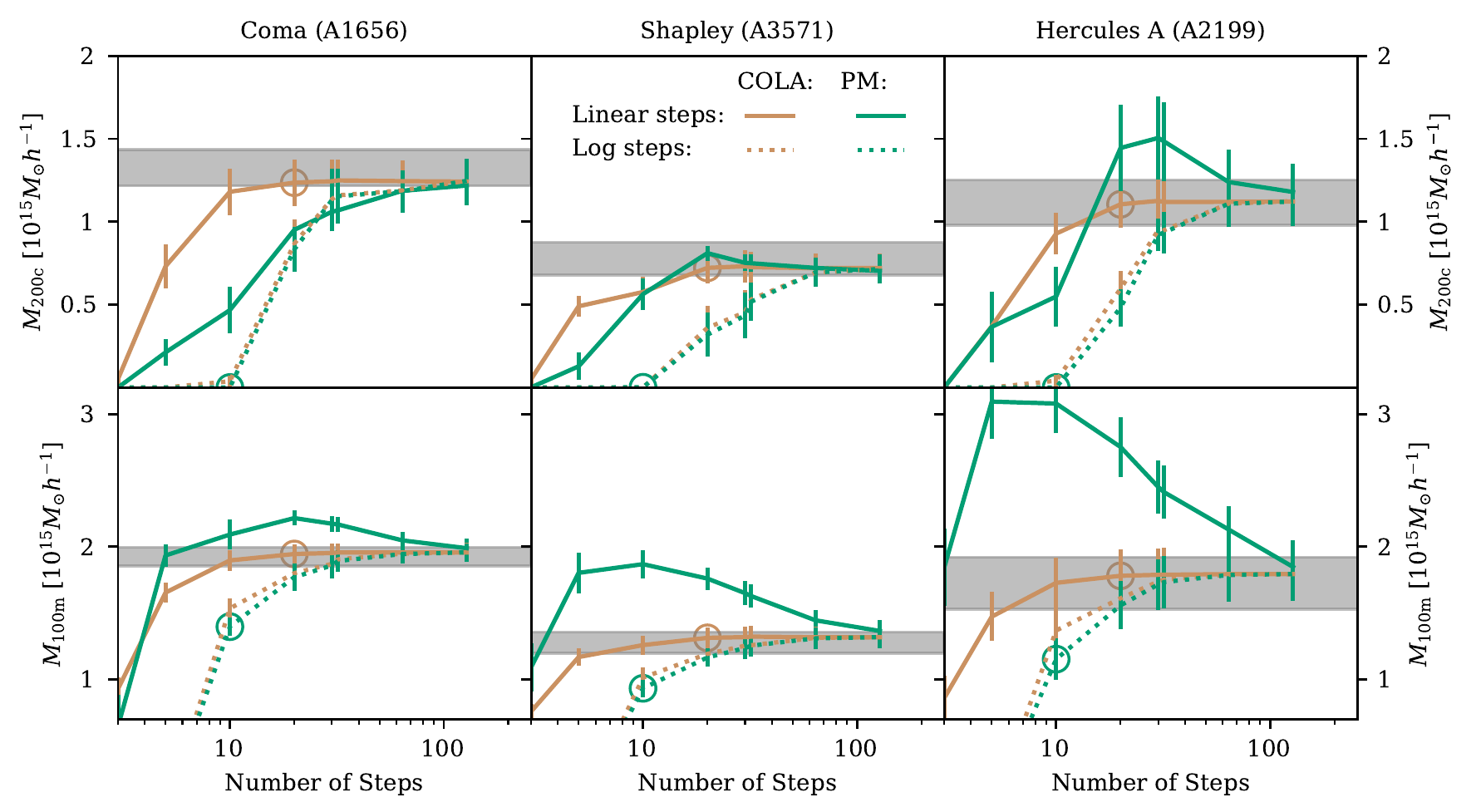}
		\caption{\label{fig:mass_convergence}Convergence of the $M_{200c}$ (upper row) and $M_{100m}$ (lower row) masses of three $\sim 10^{15}M_{\odot}h^{-1}$ clusters with \codefont{COLA} (brown lines) and \codefont{PM} (green lines) gravity solvers, starting from identical initial conditions from the~\citet{jasche2019physical} \codefont{BORG} inference. Solid lines indicate time-steps spaced linearly in scale factor, and dotted lines indicate time-steps spaced logarithmically in scale factor. Circles highlight the results for \codefont{COLA20} (brown), which is used for the chain computed in this work, and \codefont{PM10} (green) which was used for the chain generated by~\citet{jasche2019physical}. Error bars denote the standard deviation of the mean over all contributing resimulated MCMC samples. The mass of the clusters is compared to that obtained using posterior resimulation (grey region, showing the mean and the standard deviation of the mean mass over six samples from the Markov chain). The \codefont{PM10} model fails to accurately describe the mass enclosed even within the large scales measured by $M_{100m}$, but \codefont{COLA} accurately describes the mass at this scale.}
	\end{figure*}	
	
	\begin{figure}
		\centering
		\includegraphics[width=0.45\textwidth]{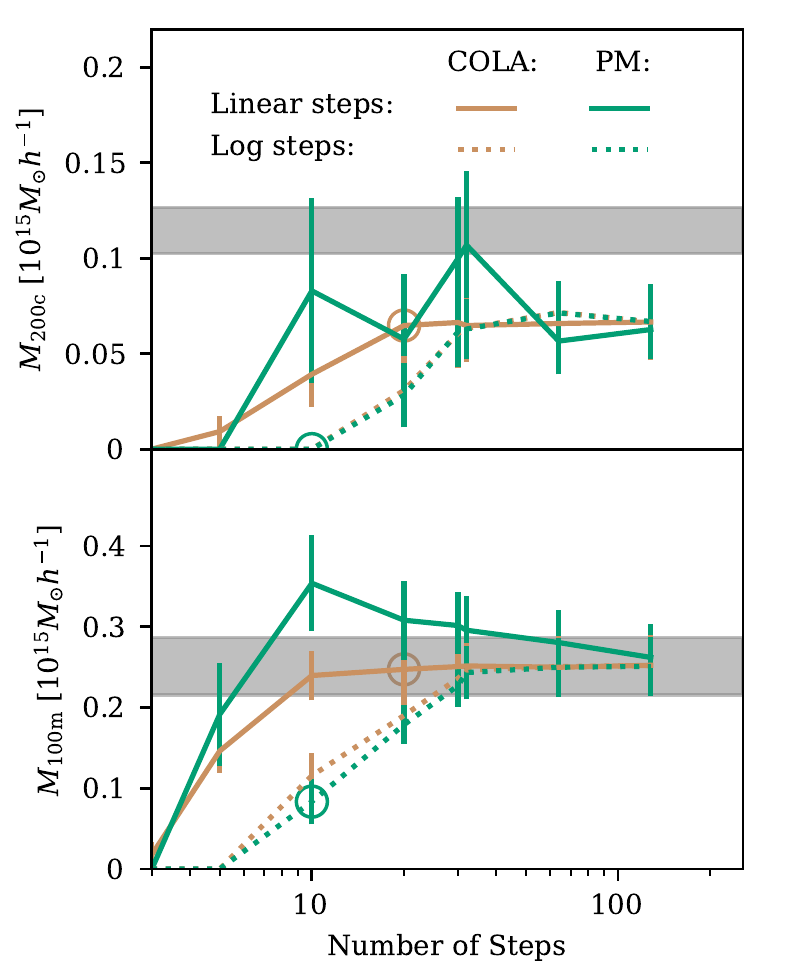}
		\caption{\label{fig:mass_convergence_other}Convergence of mass in the same simulations as Fig.~\ref{fig:mass_convergence}, but for a lower mass cluster with $M_{200c} \simeq 10^{14}M_{\odot}h^{-1}$. As for the higher mass examples, the \codefont{COLA} forward model is able to reproduce the $M_{100m}$ mass; however, due to limitations in spatial resolution it is unable to reproduce $M_{200c}$, regardless of number of timesteps. This necessitates posterior resimulation of initial conditions with \codefont{GADGET} if $M_{200c}$ masses are to be accurately recovered.}
	\end{figure}
 
	In this study, we use posterior resimulation of initial conditions obtained using field-level inference with \codefont{BORG} to estimate the masses of nearby galaxy clusters. We investigate how the accuracy of gravity solver used for field-level inference affects cluster mass estimation using posterior resimulation. Informed by these results, we select an improved gravity solver and perform a new inference of the initial conditions which achieves higher accuracy compared with those obtained by~\citet{jasche2019physical}. The latter initial conditions have been used by  \citet{desmond2022catalogues} to initialise simulations, and the resulting void properties studied. However we demonstrate that the improved accuracy obtained with our choice of gravity solver  is vital to reliably estimate both cluster and void masses.  We leave a full discussion of void properties such as their density profiles to future work.
	
	The structure of the paper is as follows. In Sec. \ref{sec:methods} we outline the methods used for field-level inference, posterior resimulation and the validation of our results. In Sec. \ref{sec:results} we determine the accuracy of the gravity solver needed for the study and show the results for cluster and void mass functions. We then produce estimates of local massive cluster masses and validate our results via comparison with existing mass estimates, as well as through internal consistency checks. We discuss the cosmological implications of the results in Sec.~\ref{sec:discussion}, including an estimate of the under-density of the local Super-Volume, and conclude in Sec.~\ref{sec:conclusions}.

	\section{Methods}
	\label{sec:methods}

	In this work, we will focus on the problem of estimating the masses of galaxy clusters. We will use field-level inference to obtain the distribution of initial conditions compatible with a galaxy catalogue, then resimulate these with greater accuracy to obtain the cluster masses themselves. The first step involves inferring the initial density field in the Lagrangian patch surrounding the galaxy clusters of interest. In particular, in order to obtain converged mass estimates via posterior resimulation in the second step, the large-scale flows in the vicinity of the cluster must be accurately inferred. This places strong requirements for the accuracy of the gravity solver used within the forward model in the field-level inference step, which we quantify in this study. 

This section is structured as follows. In Sec.~\ref{sec:borg} we describe the field-level inference framework we use, including the forward model and the dataset. In Sec. \ref{sec:mass_estimation} we discuss how posterior resimulation is used to estimate cluster masses from the field-level inference. In Sec. \ref{sec:ppt_method} we outline the techniques we use for validating our results. 
 
	\subsection{Field-level inference with BORG}
	\label{sec:borg}

    In this work, we use the \codefont{BORG}~\citep{2013MNRAS.432..894J} algorithm to perform field-level inference of galaxy cluster masses conditioned on the 2M++ galaxy catalogue~\citep{2mppPaper}. 
    \codefont{BORG} uses a Hamiltonian Markov Chain Monte Carlo (MCMC) algorithm~\citep{1987PhLB..195..216D,neal1993probabilistic} to sample the posterior distribution of possible initial density fields, $\delta_{i}^\mathrm{IC}$, assuming a $\Lambda$CDM Gaussian prior and conditioned on the observed galaxy counts, $N_i$, in a set of voxels (labelled by $i$). This posterior is given schematically by
	\begin{equation}
		P(\delta^\mathrm{IC}|N) = \frac{P(\delta^\mathrm{IC})P(N|G[\delta^\mathrm{IC}])}{P(N)},\label{eq:schematic}
	\end{equation}
	where $P(\delta^\mathrm{IC})$ is the $\Lambda$CDM prior on the initial conditions, and $P(N|G[\delta^\mathrm{IC}])$ represents the likelihood of observing a given galaxy distribution given a specific set of the initial conditions. This is dependent on a \emph{gravity solver}, $G[\delta^{\mathrm{IC}}]$, which describes the gravitational evolution which maps initial densities onto the final density field at redshift $z=0$. Additionally, a bias model is required to map the final density field into galaxy counts; while this is a crucial part of the inference, we do not represent this process in the schematic Eq.~(\ref{eq:schematic}) above, since we will not  assess the accuracy of bias modelling in this paper. We fix the bias model to that adopted by \cite{jasche2019physical}, briefly outlined in Sec.~\ref{sec:bias}, so that we can focus on the impact of the gravity solver choice on the inference accuracy.
 
 Relative to previous work presented in \citet{jasche2019physical}, we run a new MCMC inference with an improved gravity solver (see below) and the updated likelihood introduced by~\citet{porqueres2019explicit}, which is further described in Appendix \ref{app:likelihood}.

	\subsubsection{The 2M++ galaxy catalogue}
	
	The data used in the \codefont{BORG} inference in this work is identical to that used by~\citet{jasche2019physical}, known as the 2M++ galaxy catalogue~\citep{2mppPaper}. This consists of targets drawn from the 2-Micron All-Sky Survey extended source catalogue \citep[2MASS-XSC:][]{2012ApJS..199...26H}, with spectroscopic redshifts from the 2MASS Redshift Survey \citep[2MRS: ][]{2012ApJS..199...26H}. It additionally uses data from the 6-Degree Field galaxy redshift survey ~\citep[6dFGRS]{Jones:2006xy}, and the Sloan Digital Sky Survey Data Release 7~\citep[SDSS]{SDSS:2008tqn}. 
	
	Following~\citet{jasche2019physical}, we allow the \codefont{BORG} algorithm to infer galaxy bias parameters separately in several apparent and absolute magnitude bins as follows. First, the galaxies are split into two apparent magnitude bins: $K\leq 11.5$ (reaching a distance of $\simeq 200\Mpch$) and $11.5 < K \leq 12.5$ (reaching a distance of $\simeq 350\Mpch$). The vast majority of the $\simeq 70~000$ galaxies are therefore contained within the $677.7\Mpch$ simulation box. The latter does not contain any galaxies from 2MRS, due to incompleteness at fainter magnitudes. 
    These two apparent magnitude bins are further subdivided into eight absolute magnitude (denoted $M_{K}$) bins between $-25 \leq M_K \leq -21$, giving a total of 16 `catalogues' in which the bias model parameters are determined independently. The forward model  assumes that the dark matter density field is common to all 16 catalogues.

 \subsubsection{Gravity solver}

	The choice of gravity solver within the forward model is crucial to the goal of estimating the cluster masses. It is important to note that clusters themselves will not be fully resolved within the field-level inference. However, the much larger Lagrangian volume in the initial conditions can be resolved, so once the linear field on large scales is accurately inferred, the flows into a cluster are determined. This is what enables the posterior resimulation to map the initial field onto accurate cluster masses. Hence, if the gravity solver does not reconstruct these large-scale flows accurately, the inferred initial conditions, and any derived quantities, such as galaxy cluster masses, will be biased. For example, if the gravity solver underestimates the non-linear growth in high density environments, then the initial conditions will be driven towards artificially higher densities to match the galaxy catalogue. This would then result in the cluster mass being overestimated when these initial conditions are resimulated with an $N$-body code.
 
    Since the initial conditions are inferred via sampling, the gravity-solver must also be computationally efficient, in order to obtain a converged Markov chain in a reasonable time. Accuracy and speed must therefore be carefully traded off against each other to satisfy the accuracy requirements of the problem under consideration.

    In this work, we first used samples from the ~\citet{jasche2019physical} MCMC chain to test two gravity solvers: $1024^3$ particle mesh (PM)~\citep{1983MNRAS.204..891K,1997astro.ph.12217K,EASTWOOD1974342}, COmoving Lagrangian Acceleration (\codefont{COLA})~\citep{tassev2013solving}, comparing them against the adaptive-timestepping $N$-body code \codefont{GADGET2} \citep{Springel:2005mi}. We  consider the effect of spacing the time-steps linearly with scale factor, and logarithmically.  We first investigated the optimal time-stepping procedure (see Sec.~\ref{sec:timestepping} for more details), and selected a 20-step  \codefont{COLA} gravity solver (\codefont{COLA20}) with time-steps spaced linearly in scale factor to be used for our new MCMC inference. We compare our results to the 10-step particle mesh method (\codefont{PM10}) used to perform field-level inference using the same catalogue by~\citet{jasche2019physical}. 
    
    We use the solvers on a set of six initial conditions drawn from the~\citet{jasche2019physical} MCMC chain to simulate the evolution of $512^3$ particles between $z = 69$ and $z=0$ in a $677.7\Mpch$ box, for a final spatial resolution of $0.66\Mpch$. Initial conditions are inferred on a $256^3$ grid, which are oversampled to produce the $512^3$ particles used by the gravity solvers. We compare different choices for the accuracy of the gravity solver in Sec.~\ref{sec:results}. 
    
    Because we perform the above gravity solver tests with initial conditions drawn from the~\citet{jasche2019physical} MCMC chain, which was generated with the \codefont{PM10} method, the masses shown in Fig.~\ref{fig:mass_convergence} and~\ref{fig:mass_convergence_other} are not expected to be consistent with the more accurate masses that result from our new MCMC inference below. However, the convergence properties of the gravity solver are not altered by this change in the initial conditions.

	\subsubsection{Galaxy bias model}
	\label{sec:bias}

    \begin{figure*}
		\centering
		\includegraphics[width=\textwidth]{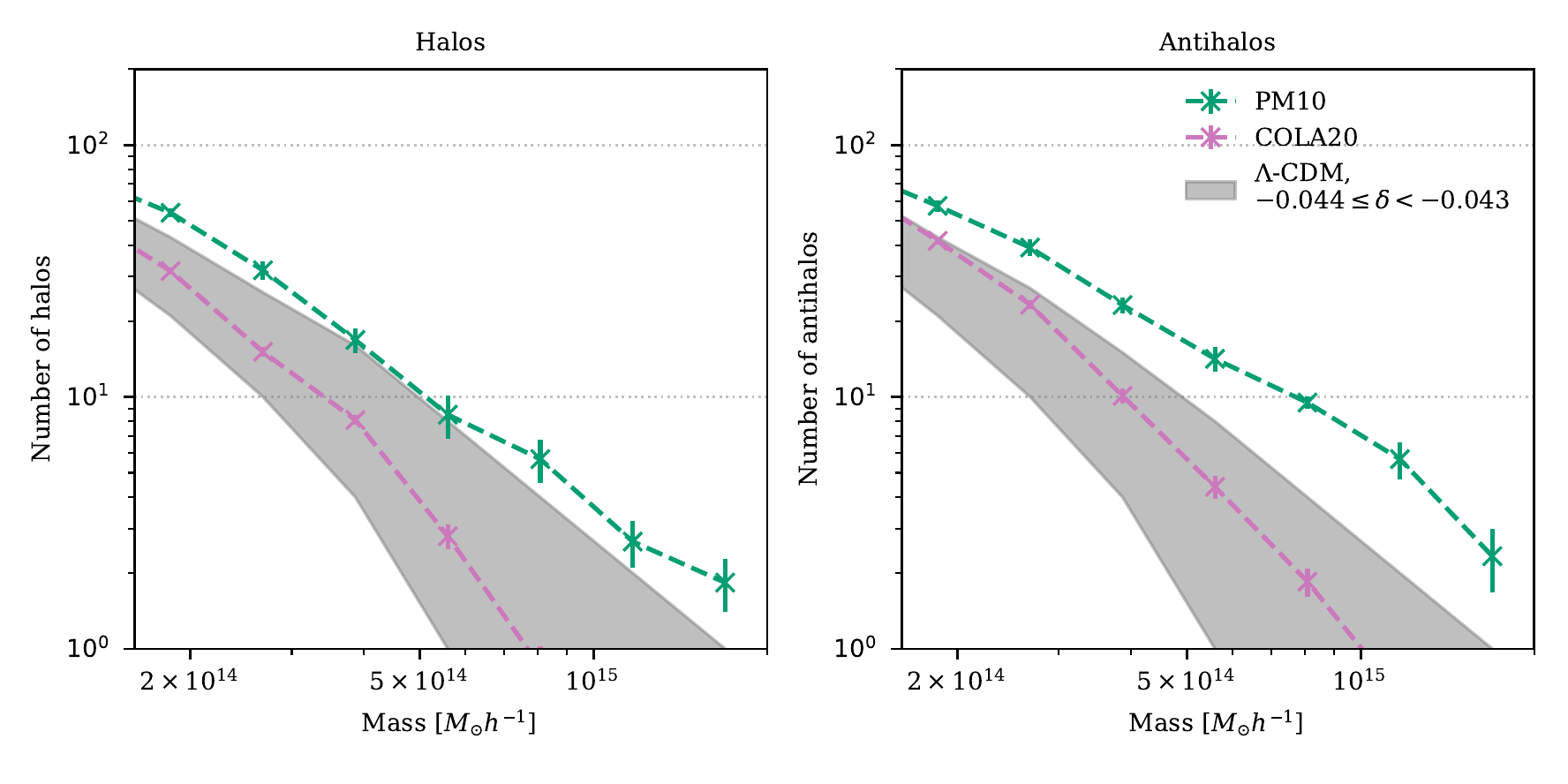}
		\caption{\label{fig:hmfsnew}Halo mass function (left) and anti-halo mass function (right) for the central $135\Mpch$ region (covering the Local Super-Volume) obtained using posterior resimulation of the new \codefont{BORG} \codefont{COLA20} field-level inference (pink lines, this work) vs the original \codefont{PM10} inference~\citep{jasche2019physical} (green lines). The shaded regions show the $95\%$ Poisson interval expected using mass functions estimated from  \codefont{GADGET} simulations with the same cosmological parameters as the \codefont{COLA20}-based inference, conditioned on regions with a density contrast that matches that of the Local Super-Volume ($\delta=-0.043\pm0.001$).}
	\end{figure*}
	
	The relationship between the \emph{final} density field and the galaxy distribution, $N_i$, is described by a \emph{galaxy bias model}. Specifically, the inference uses the~\citet{neyrinck2014halo} bias model, where the observed galaxy count $N_{i}$ in voxel $i$ is assumed to follow a Poisson distribution with mean number of galaxies $\lambda_i$. In the \citet{neyrinck2014halo} model, this mean count is related to the final density constrast in that voxel, $\delta_i$, by
	\begin{equation}
		\lambda_i(\delta_i,\bar{N},\beta,\rho_g,\epsilon_g) = S_iA_{\alpha(i)} \bar{N}(1+\delta_i)^{\beta}\exp\left(-\rho_g(1 + \delta_i)^{-\epsilon_g}\right).\label{eq:lambdai}
	\end{equation}
	The prefactor $S_i$ accounts for the selection function and survey mask in voxel $i$, constructed following the procedure of \citet{jasche2019physical}.
 The four parameters of the bias model ($\bar{N},\beta,\rho_g,\epsilon_g$) can be inferred jointly with the density field.  In practice, we infer only $\beta,\rho_g,$ and $\epsilon_g$ since we use the likelihood presented in~\citet{porqueres2019explicit}, which is insensitive to $\bar{N}$ (see Appendix~\ref{app:likelihood}). Different parameters are inferred for each of 16 galaxy catalogues (each with a different absolute and apparent magnitude range) in the 2M++ dataset.  
	
	The amplitude $A_{\alpha(i)}$ will be of particular interest in the present work. Its purpose is to account for possible unknown multiplicative, spatially-varying systematics, which may differ between each of the 16 catalogues. The $\alpha(i)$ subscript refers to the definition of the amplitudes over a separate healpix~\citep{gorski2005healpix} pixelisation of the sky with $n_{\mathrm{side}} = 4$, with each healpix pixel (healpixel) split into 10 radial bins of width $60\Mpch$ creating a set of $1,920$ regions labelled by $\alpha$. Each cubic voxel, $i$, is uniquely found in a specific region, $\alpha(i)$, which is shared by a number of voxels (see Appendix~\ref{app:likelihood} for further details). The values of $A_{\alpha}$ are marginalised over a Jeffreys prior in the likelihood, though it is necessary to reconstruct their posterior in order to perform the posterior predictive tests which we outline in Sec.~\ref{sec:ppt_method}. As a reminder that $A_{\alpha}$ can differ between catalogues, later we will refer to these values as $A_{\alpha}^{\mathrm{c}}$. 

 \subsubsection{Redshift space distortions}
 
	Redshift space distortions are treated as in \citet{jasche2019physical}: the initial density field is first evolved to $z=0$ using the gravity solver, which produces particles with known position and velocity. Then, the positions and velocities are combined to produce (physical) redshift space positions for each particle,
	\begin{equation}
		\mathbf{s} = \left(1 + \frac{a}{H(a)}\frac{\mathbf{\mathbf{v}\cdot \mathbf{r}}}{|\mathbf{r}|^2}\right)\mathbf{r},
	\end{equation}
	where $H(a)$ is the Hubble rate as a function of scale factor $a$, $\mathbf{r} = a\mathbf{x}$ is the position in physical units, $\mathbf{x}$ the comoving position, and $\mathbf{v} = \mathrm{d}\mathbf{x}/\mathrm{d}t$ is the comoving velocity. The density field in redshift space is then computed using the Cloud-in-Cell (CIC) approach on a $256^3$ grid, with a spatial resolution of $2.65\Mpch$. By applying the bias model in Eq. (\ref{eq:lambdai}) to the redshift space density field on this grid, we can compute mean galaxy counts for each voxel, which are then compared to the 2M++ galaxies on the same grid using the likelihood.

 \subsubsection{Cosmological parameters and numerical convergence}

 \begin{figure*}
		\centering
		\includegraphics[width=\textwidth]{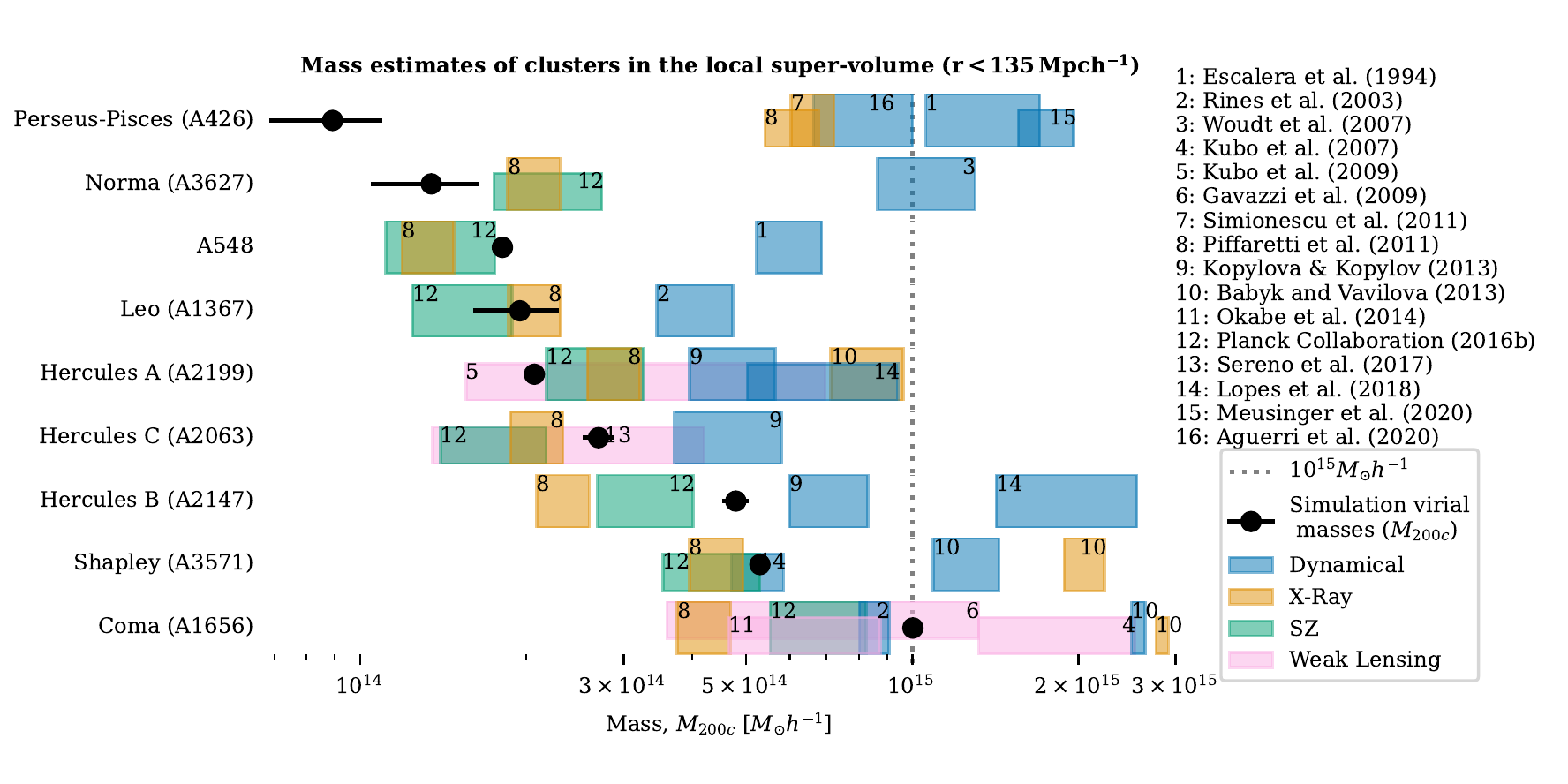}
		\caption{\label{fig:cluster_masses} Cluster mass estimates ($M_{200c}$) obtained with the resimulated \codefont{COLA20} \codefont{BORG} inference (black dots), compared with other $M_{200c}$ mass estimates for the same clusters as compiled by~\citet{stopyra2021quantifying} (non $M_{200c}$ estimates are converted $M_{200c}$ via a concentration-mass relationship). The uncertainties on the \codefont{BORG} estimates are given by the standard deviation of the distribution of halo masses associated with a given cluster, obtained from 20 resimulated MCMC samples. In most cases, the masses are consistent with other estimates. Perseus-Pisces (A426) is a notable exception, which we discuss in Sec.~\ref{sec:ppts}.}
	\end{figure*}

% Adding references from the figure to the bibliography manually:
\nocite{escalera1994structures}
\nocite{rines2003cairns}
\nocite{woudt2007norma}
\nocite{kubo2007mass}
\nocite{kubo2009sloan}
\nocite{gavazzi2009weak}
\nocite{Piffaretti_2011}
\nocite{simionescu2011baryons}
\nocite{kopylova2013investigation}
\nocite{babyk2013comparison}
\nocite{okabe2014subaru}
\nocite{Ade:2015gva}
\nocite{sereno2017psz2lens}
\nocite{lopes2018optical}
\nocite{meusinger2020galaxy}
\nocite{aguerri2020deep}
 
	For our new MCMC inference with the \texttt{COLA} solver, we assumed the Planck 2018 \citep{aghanim2020planck} cosmological parameters with lensing and baryon acoustic oscillations: $\Omega_m = 0.3111, \sigma_8=0.8102, H_0 = 67.66\,\mathrm{kms}^{-1}\mathrm{Mpc}^{-1},n_s=0.9665,\Omega_b = 0.049$. The chain was first run for 6000 MCMC steps with a 10-step \codefont{COLA} solver, after which it was switched to 20-steps and run for a further 9000 MCMC steps. The chain was converged by around 7000 MCMC steps, and we use samples from beyond 7000 steps for our results. The end product of the inference is a set of samples drawn from the posterior distribution for initial conditions consistent with the 2M++ galaxy distribution. Each sample consists of an initial density field at $z=50$ on a $256^3$ grid\footnote{Note that~\citet{jasche2019physical} used $z=69$ as their initial redshift.}, a final $256^3$ redshift-space density field at $z = 0$, and the parameters of the bias model for each of the 16 galaxy catalogues.

	\subsection{Posterior resimulation}
	\label{sec:mass_estimation}

    The second stage of our cluster mass estimation method requires resimulating many initial conditions sampled from the posterior distribution obtained using field-level inference, which can itself be a computational demanding task. However, it is not necessary to resimulate every sample from the Markov chain. One can resimulate a selection of samples from the chain and histogram the mass estimates thus obtained, assuming the estimates are approximately independent.  From these, we can compute the mean and variance of the estimated distribution. 
 
    In this work we take 20 initial conditions from the chain run in Sec.~\ref{sec:borg}, each separated by 300 MCMC steps (longer than the measured correlation length of any relevant parameter or field value). The initial conditions are generated with \codefont{genetIC}~\citep{stopyra2021genetic} from the $256^3$ grid white-noise output of \codefont{BORG}, and over-sampled with \codefont{genetIC}'s tricubic interpolation to generate $512^3$ particles with a mass resolution of $2\times 10^{11}M_{\odot}h^{-1}$, in order to reduce shot noise. They are then evolved from $z=50$ to $z=0$ with \codefont{GADGET2}~\citep{Springel:2005mi} on a $677.7\Mpch$ box, to give a set of $20$ simulations which sample the posterior distribution of the local density field.

    To find halos, we use the \codefont{AHF} halo finder~\citep{AHF}, which identifies spherical-overdensity halos in the resimulations, and except where otherwise stated we adopt the $M_{200c}$ mass definition ({\it i.e.,} the mass enclosed within a sphere whose mean density is $200$ times the critical density of the Universe).
    
    For each sample, we also perform simulations with inverted initial conditions to study the abundance of `anti-halos' as a probe of voids \citep{pontzen2016inverted}. Anti-halos are a model of voids defined in $N$-body simulations by reversing the density-contrast of the initial conditions (swapping under- and over-densities), evolving the reversed initial conditions to redshift zero, and mapping the halo particles in the resulting `anti-universe' simulation into the original simulation. Once mapped into the original simulation, they correspond to voids, with the benefit that their abundance is closely related to that of halos. Posterior resimulation is particularly well-suited to studying anti-halos, since the initial conditions are available for inversion and resimulation.

	\subsection{Posterior predictive tests}
	\label{sec:ppt_method}

 \begin{figure*}
		\centering
		\includegraphics[width=\textwidth]{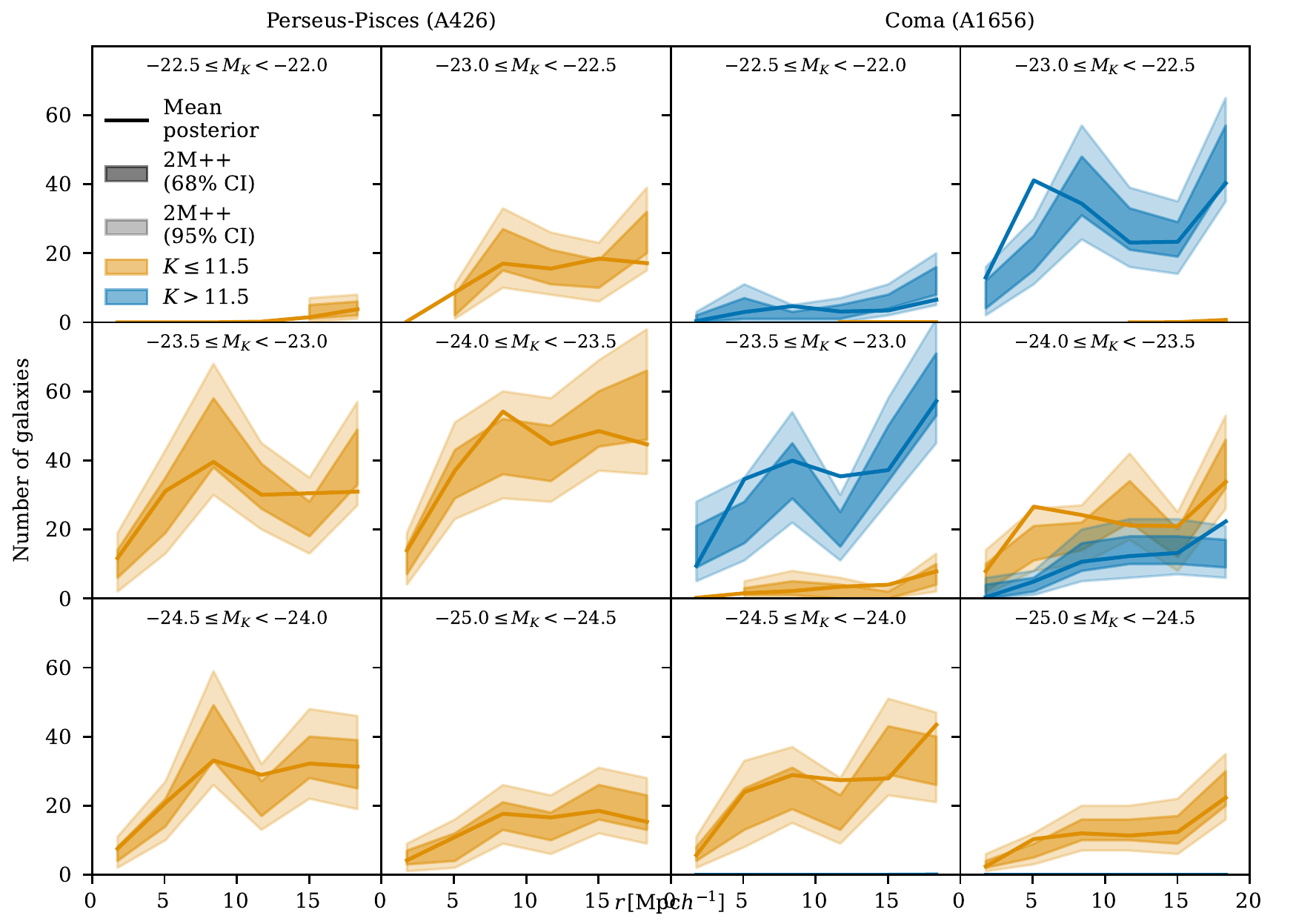}
		\caption{\label{fig:ppt} Posterior predictive tests for the galaxy counts in radial $4\Mpch$-wide shells around two clusters in the Local Super-Volume: Perseus-Pisces (left panels) and Coma (right panels). Solid lines show the predicted mean counts from the posterior distribution for each absolute magnitude bin, while shaded regions show the $68\%$ and $95\%$ credible intervals computed by bootstrapping the sum of all voxels in each shell for the 2M++ galaxies. The $K \leq 11.5$ catalogue is shown in orange, while the $K > 11.5$ catalogue is shown in blue. Note that Perseus-Pisces entirely lacks $K > 11.5$ catalogue data due to being in the 2MRS portion of the sky. Despite its apparently underestimated mass in Fig.~\ref{fig:cluster_masses}, the posterior predictive tests pass in all magnitude bins.}
	\end{figure*}

	To verify the accuracy of the inferred initial conditions, we perform posterior predictive tests to compare the posterior-predicted galaxy counts to those found in the 2M++ galaxy catalogue. For this task, we require the posterior distribution on the expected number of galaxies in the $i$th voxel, $\lambda_i$, given observed 2M++ galaxy counts. Obtaining this is complicated by the fact that (see Appendix~\ref{app:likelihood}) the likelihood used in this work marginalises over the amplitudes $A_{\alpha}^{\mathrm{c}}$ in Eq.~(\ref{eq:lambdai}), and hence this posterior is not explicitly provided by the \codefont{BORG} MCMC chain. However, the required posterior can be constructed from the MCMC chain, since it provides samples from the posterior on $\bar{\lambda}_i^{\mathrm{c}} \equiv  \lambda_i^{\mathrm{c}} / A_{\alpha(i)}^\mathrm{c}$. 
     We find (see Appendix~\ref{app:ppt_derivation} for details) that the expectation value of $A_{\alpha}^\mathrm{c}$ for catalogue $\mathrm{c}$ is given by 
	\begin{equation}
		E(A_{\alpha}^\mathrm{c}|N^\mathrm{c}_{1 \cdots I}) \simeq \frac{N_{\mathrm{tot},\alpha}^\mathrm{c}}{S} \sum_{s=1}^{S} \frac{1}{\bar{\lambda}^\mathrm{c}_{\mathrm{tot},\alpha,s}}\, ,\label{eq:Aalpha}
	\end{equation}
	where $i=\{1 \cdots I\}$ are assumed to be the unmasked voxels in the healpixel $\alpha$, $s$ indexes $S$ samples from the posterior, $N_{\mathrm{tot},\alpha}^{\mathrm{c}} = \sum_{i=1}^{I} N_i^{\mathrm{c}}$, and $\bar{\lambda}^{\mathrm{c}}_{\mathrm{tot},\alpha,s} = \sum_{i=1}^I\bar{\lambda}^{\mathrm{c}}_{i,s}$. Furthermore, the posterior distribution $P(\lambda_i^{\mathrm{c}}|N^\mathrm{c}_{1 \cdots I}) = P(A_{\alpha}^\mathrm{c}\bar{\lambda}^\mathrm{c}_i|N^\mathrm{c}_{1 \cdots I})$ has expectation value 

\begin{equation}
    E(A_{\alpha}^\mathrm{c}\bar{\lambda}^\mathrm{c}_i|N^\mathrm{c}_{1 \cdots I}) \simeq 
 \frac{N_{\mathrm{tot},\alpha}^\mathrm{c}}{S} \sum_{s=1}^{S} \frac{\bar{\lambda}^\mathrm{c}_{i,\alpha,s}}{\bar{\lambda}^\mathrm{c}_{\mathrm{tot},\alpha,s}}\, .\label{eq:Elambdai}
\end{equation}
 
Given that the setup using multiple amplitudes and catalogues makes the correlation structure of the inferred field and hence its variance cumbersome to calculate analytically, we proceed using a bootstrap estimate of the uncertainty on the 2M++ data. In particular, for each of the $16$ galaxy catalogues, we bootstrap the 2M++ galaxy counts for the voxels in each spherical shell to obtain an estimate of the uncertainty on the total number of galaxies in that shell.

	\section{Results}
	\label{sec:results}

    \begin{figure*}
		\includegraphics[width=\textwidth]{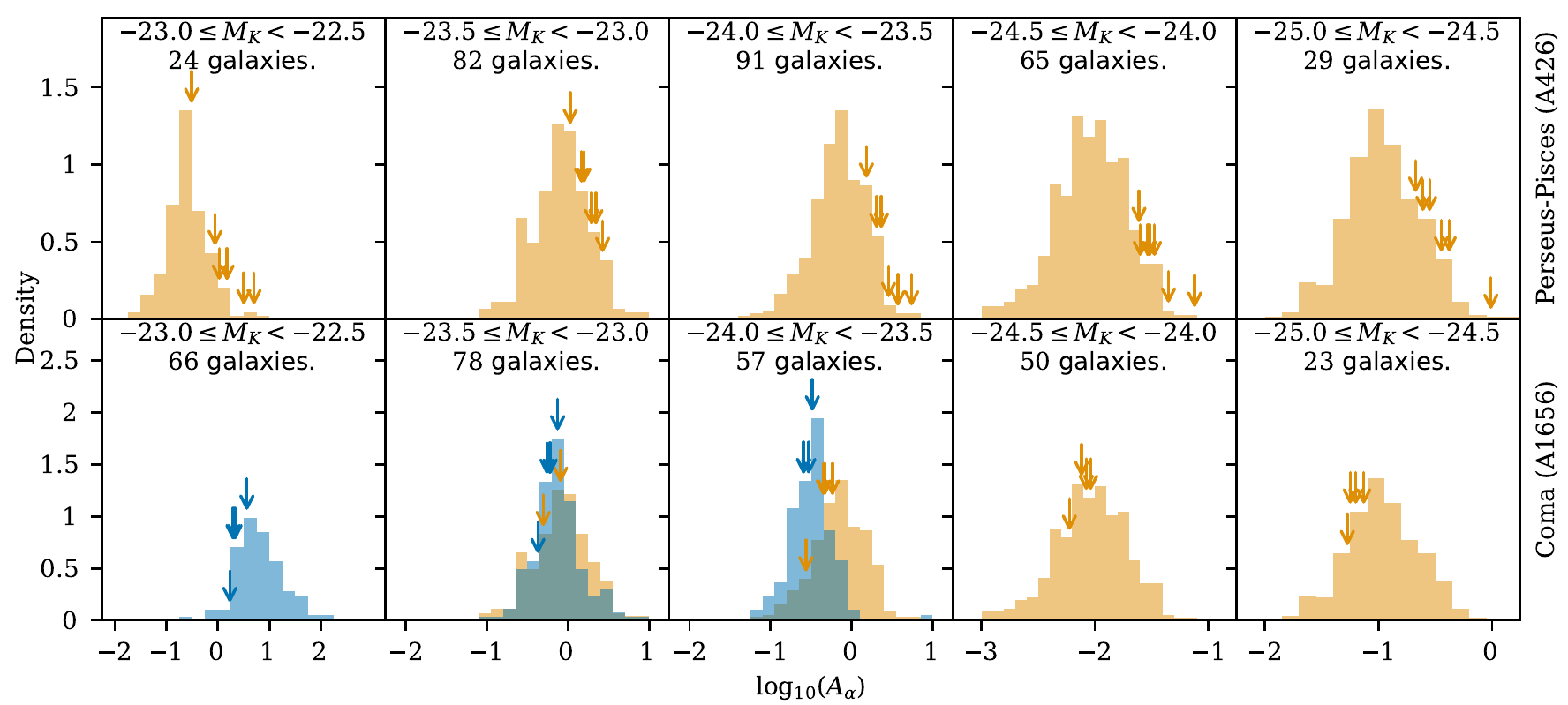}
		\caption{\label{fig:amplitudes}Distribution across the full sky of expected amplitudes, $A_{\alpha}$, given by Eq.~(\ref{eq:Aalpha}) for each of five magnitude bins (histograms), with arrows indicating the amplitudes for the healpixels within $10\Mpch$ of the centre of the two clusters Perseus-Pisces (top) and Coma (bottom). Orange again indicates the $K \leq 11.5$ catalogue, while blue indicates the $11.5 < K \leq 12.5$ catalogue. The number of 2M++ galaxies within $10\Mpch$ of the cluster centre which lie in each absolute magnitude bin are specified. The healpixel amplitudes for Perseus-Pisces consistently lie in the high-amplitude tail of the distribution. Note that the distance to Perseus-Pisces ($54.6\Mpch$) places it close to the boundary between two different $60\Mpch$-wide healpix shells. It therefore  receives contributions from many more healpix regions than Coma.}
	\end{figure*}

	This section is structured as follows. In Sec.~\ref{sec:timestepping}, by considering the convergence of mass estimates as a function of physical scale, we show that the \codefont{COLA20} gravity solver is adequate for obtaining reliable mass estimates for the highest mass clusters. In Sec.~\ref{sec:mass-functions} we present the results for mass functions obtained using \codefont{COLA20}, and compare it to the previous state-of-the-art \codefont{PM10} inference. In Sec.~\ref{sec:individual-cluster-masses} we discuss how individual clusters with masses between $10^{14}$ and $10^{15} M_{\odot}\,h^{-1}$ can be identified within the posterior resimulations, comparing the results with a collection of mass estimates from the literature.  Finally, we present the results of posterior predictive tests for the galaxy counts in these clusters, noting some possible indications of remaining systematic uncertainties (Sec.~\ref{sec:ppts}).

    \subsection{Choice of gravity solver}
	\label{sec:timestepping}

    We begin by establishing the accuracy needed for the gravity solver within the field-level inference in order to obtain reliable cluster masses. This is accomplished by testing which gravity solvers can predict converged masses relative to a \codefont{GADGET2} resimulation, starting from the same initial conditions. We used initial conditions from the \codefont{PM10} MCMC chain~\citet{jasche2019physical} at $z=69$ and evolved them to $z=0$ using a variety of solvers with different, fixed numbers of time-steps. We identified massive clusters in the simulations by searching for the largest halo within a $20\Mpch$ radius of the known position of a cluster in the Abell catalogue~\citep{abell1989catalog}. We examined two different definitions of mass ($M_{200c}$, the mass contained within a sphere of density $200$ times the critical density, and $M_{100m}$, the mass contained within a sphere of density 100 times the mean density of the Universe). This allowed us to examine how the gravity solvers perform at different scales: $M_{200c}$ virial radii are typically half that of $M_{100m}$ virial radii. Convergence to the correct mass for a given set of initial conditions indicates that the solver is consistent with $N$-body simulations at a particular scale.
		
	We show the results of the time-step tests for several different clusters in the~\citet{jasche2019physical} MCMC chain in Fig.~\ref{fig:mass_convergence} for high mass, $\sim 10^{15}M_{\odot}h^{-1}$ clusters.\footnote{Note that because these tests were done with initial conditions from the~\citet{jasche2019physical} MCMC chain which used a \codefont{PM10} gravity solver, the masses are not expected to be consistent with the masses found for these clusters in Fig.~\ref{fig:cluster_masses}.} In each case, we consider a range of step numbers between $3$ and $128$, and we also consider linear or logarithmic spacing of these steps in scale factor between $z=69$ and $z=0$. Grey bands show the standard deviation of the mean mass for these halos over the samples taken from the posterior, while the lines show the mean mass over all contributing MCMC samples (error bar is the standard deviation of the mean). 

    At this mass scale, all the gravity solvers we considered converged to the same mass as \codefont{GADGET} simulations when the number of time-steps is increased, representing the increased accuracy (but also increased computational cost) that comes with using more time-steps. For the \codefont{COLA} model, $10$--$20$ time-steps was found to be sufficient to reproduce the mass of the largest clusters ($\sim 10^{15}M_{\odot}h^{-1}$) at a level consistent with \codefont{GADGET} (see Fig.~\ref{fig:mass_convergence}).

    We repeated our test for lower-mass clusters (${\sim 10^{14}M_{\odot}h^{-1}}$), with an example being shown in Fig.~\ref{fig:mass_convergence_other}. In these cases, increasing the timestepping accuracy does not result in convergence to the \codefont{GADGET} result for $M_{200c}$. This implies that the primary limitation on the accuracy of the gravity solver is spatial rather than temporal resolution on these scales. The grid scale is $2.65\,\Mpch$, considerably larger than the virial radius $R_{200c} \simeq 0.75\,\Mpch$ for this definition.   However, convergence is still achieved for the $M_{100m}$ masses, and at this mass scale $R_{100m} \simeq 1.4\,\Mpch$. This implies that the overall mass in the cluster environment can be correctly reproduced with an approximate gravity solver, provided one does not extrapolate too far below the grid scale.\footnote{Note that recovering the $M_{200\mathrm{c}}$ mass of halos is generally a more stringent test than, for example, reproducing the Friend-of-Friends halo mass function with linking length $l=0.2\Mpch$, which can be done with approximate solvers such as \codefont{COLA} or \codefont{FASTPM}~\citep{feng2016fastpm} at the percent level~\citep{10.1093/mnras/stw797}.}

    Choosing a specific gravity solver, timestep number, and timestep spacing implies a trade-off between accuracy and runtime. Ideally, one would choose a timestep that achieves convergence for spatially-resolved quantities. Based on the results discussed above, we chose \codefont{COLA} with 20 linearly-spaced steps as the best trade-off for the spatial resolution of the inference here. While we did not consider other integrators such as \codefont{FASTPM} in this work, recent work by~\citet{2023arXiv230109655L} suggests that the incorporation of Lagrangian perturbation theory information into an integrator allows for optimal use of time-steps. We therefore expect that \codefont{FASTPM} should perform similarly well to \codefont{COLA} for this purpose.

    We emphasise that, while spatial resolution prevents $M_{200c}$ convergence for low-mass clusters ($\sim 10^{14}\Mpch$) within the \codefont{BORG} gravity solver,  resimulations with \codefont{GADGET} starting from posterior samples will overcome this limitation. Based on the understanding that $M_{200c}$ is dictated by the larger-scale flows that are resolved, it is therefore legitimate to test whether the resimulated halo mass functions agree with expectations.

 \subsection{Mass functions}\label{sec:mass-functions}

   	 In Fig.~\ref{fig:hmfsnew}, we show the halo and anti-halo $M_{200c}$ mass functions inferred when using posterior resimulation applied to samples from \codefont{BORG} with the \codefont{COLA20} gravity solver, and compare this with the same results using the previous \codefont{PM10}-based inference.     

    Fig.~\ref{fig:hmfsnew} shows that the \codefont{COLA20} mass functions are in agreement with those obtained from regions of similar underdensity in unconstrained simulations (shown by the blue shaded region), and are therefore compatible with $\Lambda$CDM expectations. Conversely, when using the \codefont{PM10} gravity solver from~\citet{jasche2019physical}, the resimulated mass functions overpredict the halo and anti-halo abundances. To obtain `similar underdensity' regions in unconstrained simulations, we first computed the mean density contrast of the central $135\Mpch$ region over all $20$ MCMC samples from the \codefont{COLA20} chain, which gave $\delta = -0.043\pm 0.001$ (standard error of the mean). We then select $135\Mpch$ spheres centered randomly within a set of unconstrained simulations with cosmological parameters matching that of the \texttt{BORG} inference, and retain those spheres whose density contrast lies within one standard deviation of this mean ($-0.044 < \delta< -0.042$).  	

    These results show that insufficient integration accuracy leads the sampler to push linear over- and under-densities to exaggerated values in compensation; when re-simulated, this leads to unphysically high mass clusters and anti-halos. In more detail, the \codefont{PM10} solver presented by~\citet{jasche2019physical} has insufficient time-steps at low redshift to properly resolve the final-stage collapse of even high-mass halos. As a result, it underestimates the true density at the core of massive halos (Fig.~\ref{fig:mass_convergence}); this causes \codefont{BORG} to overestimate the initial conditions in order to infer the correct final density field. This results in inflated cluster masses when the initial conditions are resimulated with more accurate $N$-body solvers. This indirectly affects the masses of the anti-halos as well, since consistency with the larger-scale under-density of the Local Super-Volume requires the extra mass in these clusters to be taken from surrounding regions, resulting in an excess of anti-halos.	
         
     These results explain the excess of clusters found by \codefont{SIBELIUS-DARK}~\citep{mcalpine2022sibelius} and~\citet{hutt2022effect}, as well the excess of anti-halos relative to $\Lambda$CDM noted in~\citet{desmond2022catalogues}; all of these works used the \citet{jasche2019physical} inference based on 2M++. Our \codefont{COLA20}-based result demonstrates that the excess of anti-halos and halos is not a real effect, and disappears when a more accurate gravity solver is used for the inference. The requirements for reconstructing {\it individual} cluster masses, however, may be even more stringent than the requirements for the mass function, and we now turn to this issue.

	\subsection{Individual cluster masses}\label{sec:individual-cluster-masses}
	\label{sec:hmf}

  In the resimulations based on \codefont{COLA20}, we again identify clusters by searching for the largest halo within a $20\Mpch$ radius, as we previously applied to the older \codefont{PM10}-based resimulations (Sec.~\ref{sec:timestepping}). This leads to an unambiguous identification of the relevant halo for clusters of masses approaching $~10^{15}M_{\odot}h^{-1}$ which are well-constrained by the field-level inference. There are nine such cases, and we identify the clusters as Perseus Pisces (A426), Hercules B (A2147), Coma (A1656), Norma (A3627), Shapley (A3571), A548, Hercules A (A2199), Hercules C (A2063), and Leo (A1367). The mean mass of each cluster using the posterior distribution is obtained by averaging the halo masses for all its counterparts across all 20 resimulations.

 To compare these $M_{200c}$ mass estimates to known data, we make use of previously collected mass estimates for nine nearby massive clusters. The estimates come from dynamical, X-ray, SZ and weak lensing techniques; full details of the estimates are discussed in~\citet{stopyra2021quantifying}. 
	These mass estimates are shown in Fig.~\ref{fig:cluster_masses} and compared to our resimulation estimates using the new \codefont{COLA20}-based field-level inference. 
 In most cases, these new results are consistent with existing mass estimates. Perseus-Pisces (A426) is a notable exception, which we will return to in Sec.~\ref{sec:ppts}.
 
 For several of these clusters, $M_{200c}$ is in the order of $10^{14}\,M_{\odot}h^{-1}$. As previously discussed in relation to Fig.~\ref{fig:mass_convergence_other}, the gravity solver used within the inference is prevented by its spatial resolution from directly predicting such masses; the resimulations, however, remove this limitation.

	\subsection{Posterior predictive tests}
	\label{sec:ppts}

 So far, we have shown that using \codefont{COLA20} as the gravity integrator within the \codefont{BORG} pipeline, then resimulating to obtain a final cluster (or anti-halo) catalogue, gives rise to  $M_{200c}$ cluster and anti-halo mass functions in good agreement with $\Lambda$CDM expectations.  We have further shown that the mass estimates for individual clusters obtained in this way are in most cases in agreement with independent observational estimates. 
	 
    We next computed the posterior predictive tests outlined in Sec.~\ref{sec:ppt_method} for the nine individual massive clusters discussed in the previous section. Since the majority of the clusters pass the posterior predictive test we show in Fig.~\ref{fig:ppt} just two illustrative examples: Coma and Perseus-Pisces, which contain similar numbers of galaxies. Posterior predictive tests for the other seven clusters are shown in Appendix \ref{app:ppts_other}.
    The dark and light shaded regions of Fig.~\ref{fig:ppt} indicate $68\%$ and $95\%$ credible intervals for the 2M++ galaxy counts in radial shells around the centres of each cluster, estimated using bootstrap. The solid lines show the mean of the posterior predicted galaxy counts in the same shells computed using Eq.~(\eqref{eq:Elambdai}). Orange and blue colours respectively indicate the $K\leq 11.5$ and $K>11.5$ catalogues.

    In the case of Perseus-Pisces, while the mass appears substantially underestimated relative to constraints in the literature (as shown in Fig.~\ref{fig:cluster_masses}), the posterior-predicted number of galaxies is consistent with the available data. This raises the question of how the mass can be so low compared to Coma which has a similar number of observed galaxies but an order of magnitude higher inferred mass. The connection from the inferred density field to predicted number counts is dictated by a global bias model per catalogue, but it is additionally locally modulated in each healpix pixel by an amplitude $A^\mathrm{c}_{\alpha}$. We therefore investigated the expectation value of $A^\mathrm{c}_{\alpha}$ for the healpixel in the $10\,\Mpch$ surrounding Coma and Perseus-Pisces to test whether these amplitudes can account for the differing results. 

    Fig.~\ref{fig:amplitudes} shows, in absolute magnitude bins (left to right) for Perseus-Pisces (top) and Coma (bottom), the distribution of the expectation values for healpixel amplitudes across the entire Local Super-Volume. These are computed using Eq.~(\ref{eq:Aalpha}), and $K\leq11.5$ and $K>11.5$ results are respectively shown as orange and blue histograms. Vertical arrows show the amplitudes for healpixels within $10\Mpch$ of the centre of each cluster. First, we note that Perseus-Pisces has no galaxies in the $K>11.5$ catalogues (and therefore these histograms are not shown). Second, in the bright ($K\leq 11.5$) galaxies, the amplitudes inferred around Perseus-Pisces are preferentially in the high-end tail. This is indicative of a potential systematic error, allowing an unphysically small overdensity to generate a large number of galaxies. However, obtaining further insight into this result requires a detailed study of the interaction between the local amplitudes and the global bias models, or even a revision of the construction of the original catalogue, which is beyond the scope of the present work. We also note that Perseus-Pisces appears to lack SZ or weak lensing mass estimates in the literature, and that compiled dynamical mass estimates for other clusters are nearly all overestimates relative to other methods. In summary, there are strong indications that our posterior resimulation mass inference for Perseus-Pisces is biased, but we also cannot rule out the possibility that mass estimates using other methods for Perseus-Pisces may require revision.

	\section{Discussion}
	\label{sec:discussion}

	In a previous work, we showed that the number of clusters and anti-halos in the Local Super-Volume with masses above $10^{15}M_{\odot}h^{-1}$ is a powerful test of consistency with $\Lambda$CDM~\citep{stopyra2021quantifying}. Related cosmological tests include assessments of the rarity of structures such as the Sloan Great Wall, and the Shapley supercluster~\citep{nichol2006effect,sheth2011unusual}. While the 2M++ catalogue does not have sufficient signal-to-noise to directly assess the rarity of the Sloan Great Wall, we find that the number of massive clusters in Fig.~\ref{fig:cluster_masses}, and the mass function in Fig.~\ref{fig:hmfsnew} appear compatible with $\Lambda$CDM, with masses broadly consisted with X-ray and SZ estimates. Our results show that the number of massive clusters and anti-halos in the Local Super-Volume is compatible with theoretical expectations from a $\Lambda$CDM universe. 
	
	The question of whether the Local Super-Volume lies at the centre of a large Local Void has attracted much attention in the literature~\citep{xie2014local,frith2003local,shanks2019local}. More recently, it has been suggested that a large scale underdensity might be responsible for the Hubble tension~\citep{shanks2019gaia}. In particular, the 2MASS survey has been claimed to provide evidence for such a large-scale underdensity~\citep{frith2003local,shanks2019local}. Using field-level inference with \codefont{BORG} -- which includes 2MASS data as part of the 2M++ catalogue -- we are able to directly probe the dark matter density field and thus infer the size of the large-scale under-density. We find that the Local Super-Volume has a density contrast of $\delta = -0.043\pm 0.001$ out to a radius of $135\Mpch$. While this means that the Local Super-Volume is underdense, it is not sufficiently underdense to have a significant impact on measurements of the Hubble rate~\citep{wu2017sample}, so does not help by itself to alleviate the Hubble tension.	
	
	\section{Conclusions}\label{sec:conclusions}
	
	In this work, we assessed different gravity solvers that can be used for field-level inference in the context of the \codefont{BORG} algorithm followed by posterior resimulation. We showed that replacing the 10-step \codefont{PM} solver used in the~\citet{jasche2019physical} with a 20-step \codefont{COLA} solver corrects the overabundance of massive halos and voids previously noted by \cite{mcalpine2022sibelius,hutt2022effect,desmond2022catalogues}. 
	
	We used the 20-step \codefont{COLA} solver to construct a new \codefont{BORG} inference from the 2M++ catalogue, with updated cosmological parameters and likelihood. The new inference, combined with posterior resimulation, leads to halo and void mass functions which are consistent with $\Lambda$CDM expectations, and the masses of individual massive local clusters are broadly consistent with other estimates in the literature. An exception to this, however, was the Perseus-Pisces cluster, which appeared to have a significantly lower mass than expected from its galaxy counts and from other estimates of its mass in the literature. We discussed possible reasons for this in Sec. \ref{sec:ppts}.
	
	Our results show that the use of field-level inference for extracting information on non-linear scales requires careful control of the accuracy of the forward modelling used. We have shown how to validate the forward modelling for the accurate estimation of cluster masses using posterior resimulation. In the process, we have created a new inference of the intial conditions consistent with the Local Universe and provided constraints on the local underdensity, and the number of massive nearby clusters. In future work we will apply this inference to constructing an anti-halo void catalogue of the Local Super-Volume.

	\section*{Acknowledgments}
 We thank Daniel Mortlock, Harry Desmond, Stuart McAlpine, and Metin Ata for useful discussions regarding this work. This project has received funding from the European Research Council (ERC) under the European Union’s Horizon 2020 research and innovation programmes (grant agreement no. 101018897 CosmicExplorer and 818085 GMGalaxies). This work has been enabled by support from the research project grant ‘Understanding the Dynamic Universe’ funded by the Knut and Alice Wallenberg Foundation under Dnr KAW 2018.0067. SS and HVP are additionally supported by the Göran Gustafsson Foundation for Research in Natural Sciences and Medicine. HVP and JJ acknowledge the hospitality of the Aspen Center for Physics, which is supported by National Science Foundation grant PHY-1607611. The participation of HVP and JJ at the Aspen Center for Physics
was supported by the Simons Foundation. JJ acknowledges support by the Swedish Research Council (VR) under the project 2020-05143 -- ``Deciphering the Dynamics of Cosmic Structure" and by the 
Simons Collaboration on “Learning the Universe”. This work was partially enabled by the UCL Cosmoparticle Initiative. The computations/data handling were enabled by resources provided by the Swedish National Infrastructure for Computing (SNIC) at Link\"{o}ping University, partially funded by the Swedish Research Council through grant agreement no. 2018-05973. This research also utilized the Sunrise HPC facility supported by the Technical Division at the Department of Physics, Stockholm University. This work was carried out within the Aquila Consortium\footnote{\url{https://aquila-consortium.org}}.

	\section*{Author Contributions}
 The first four authors contributed roughly equally to this work, with differing areas of contribution. GL developed and implemented the \codefont{tCOLA} gravity solver used by \codefont{BORG}, and set up the \codefont{COLA20} chain used in this analysis. We outline the different contributions below using key-words based on the CRediT (Contribution Roles Taxonomy) system.
 
{\bf SS}: data curation; investigation; formal analysis; software; visualisation; writing -- original draft preparation.

{\bf HVP}: conceptualisation; methodology; validation and interpretation; writing (original draft; review and editing); funding acquisition. 

{\bf AP}: conceptualisation; methodology; validation and interpretation; writing (original draft; review and editing).

{\bf JJ}: methodology; data curation; resources; software; validation and interpretation; writing – review

{\bf GL}: data curation; resources; software; writing -- review.

	\section*{Data availability}
	The data underlying this article will be shared on reasonable request to the corresponding author.
	
	\appendix

	\section{Likelihood}
	\label{app:likelihood}
	The new \codefont{BORG} inference we used in this paper uses an updated likelihood  compared to the~\citet{jasche2019physical} inference, first outlined by~\citet{porqueres2019explicit}. The aim of this approach is to make the likelihood robust to unknown systematics in the underlying data. 
	
	To achieve this, the sky is first subdivided into 192 healpix pixels~\citep{gorski2005healpix}, with $n_{\mathrm{side}}=4$. The extension of these healpixels into three dimensions is then split into 10 radial bins of depth $60\Mpch$ each, giving 1920 regions on the sky which we denote \emph{patches}. Note that this is a separate pixelisation to the $256^3$ cubic voxels into which we divide the simulation. Each voxel, which we label with Roman letters, $i$, lies within a given healpix patch (labelled by Greek letters, $\alpha$), and there is a unique mapping $\alpha(i)$ from voxels to the healpix patch which contains them.
	
	All the voxels which lie in a given patch have a shared amplitude, $A_{\alpha}^{\mathrm{c}}$, for catalogue $\mathrm{c}$, which is regarded as a nuisance parameter and marginalised over in the likelihood. Once the gravity solver is specified, Eq.~(\ref{eq:lambdai}) gives the expected number of galaxies for catalogue $\mathrm{c}$ in voxel $i$, $\lambda_i^{\mathrm{c}}$, as a function of the final density field, $\mathbf{\delta}=G(\mathbf{\delta}^\mathrm{IC})$. We define $\bar{\lambda}_{i}^{\mathrm{c}}=\lambda_i^{\mathrm{c}}/A_{\alpha(i)}^{\mathrm{c}}$, where $A_{\alpha(i)}^{\mathrm{c}}$ is the systematics amplitude parameter corresponding to the patch containing voxel $i$. We can then write a Poisson  likelihood for the number of galaxies in catalogue $\mathrm{c}$, $N_i^{\mathrm{c}}$, that lie in voxels $i\in\{1,\ldots,I\}$ in patch $\alpha$ as
	
	\begin{align}
		P(N_{1\ldots I}^{\mathrm{c}}|\bar{\lambda}_{1\ldots I}^{\mathrm{c}},A_{\alpha}^{\mathrm{c}}) = \prod_{i=1}^{I}\frac{(A_{\alpha}^{\mathrm{c}}\bar{\lambda}_{i}^{\mathrm{c}})^{N_{i}^{\mathrm{c}}}}{N_i^{\mathrm{c}}!}e^{-A_{\alpha}^{\mathrm{c}}\bar{\lambda}_{i}^{\mathrm{c}}}. \label{eq:Poisson_like}
	\end{align}
 \citet{porqueres2019explicit} chose to marginalise over the amplitudes $A_{\alpha}^{\mathrm{c}}$ to create a new `robust' likelihood that reduces sensitivity to spatially-varying systematics in the data,
	\begin{equation}
		P(N_{1\ldots I}^{\mathrm{c}}|\bar{\lambda}_{1\ldots I}^{\mathrm{c}}) = \int\mathrm{d}A_{\alpha}^{\mathrm{c}}P(A_{\alpha}^{\mathrm{c}})P(N_{1\ldots I}^{\mathrm{c}}|\bar{\lambda}_{1\ldots I}^{\mathrm{c}},A_{\alpha}^{\mathrm{c}})\, .
	\end{equation}
	They chose a Jeffreys prior $P(A_{\alpha}^{\mathrm{c}})=\kappa_{\alpha}/A_{\alpha}^{\mathrm{c}}$, with normalisation constant $\kappa_{\alpha}$. This yields the marginalised likelihood
	\begin{equation}
		P(N_{1\ldots I}^{\mathrm{c}}|\bar{\lambda}_{1\ldots I}^{\mathrm{c}}) \propto 
  \prod_{i=1}^I\left(\frac{\bar{\lambda}_i^{\mathrm{c}}}{
\sum_{j=1}^{I}\bar{\lambda}_j^{\mathrm{c}}}\right)^{N_i^{\mathrm{c}}}.\label{eq:likelihood}
	\end{equation}
	Note that we assume all catalogues $\mathrm{c}$ are independent except via their dependence on the final density $\delta_{i}$ used to compute $\bar{\lambda}_i^{\mathrm{c}}$ in Eq.~(\ref{eq:lambdai}). The likelihood for the number of galaxies in all catalogues and voxels is therefore given by the product of Eq.~(\ref{eq:likelihood}) over all catalogues, $\mathrm{c}$, and all patches, $\alpha$.
	
	\section{Posterior Predictive Tests with the Robust Likelihood}
	\label{app:ppt_derivation}
	
	The fact that the amplitudes in Eq.~(\ref{eq:lambdai}) are marginalised over in the robust likelihood presented in Appendix~\ref{app:likelihood} makes it more difficult to interrogate the inferred model using posterior predictive testing. We wish to compare the posterior-predicted number of galaxies to that actually found in the 2M++ catalogue in specific regions of interest; this requires a posterior for the amplitudes, $A_{\alpha}^{\mathrm{c}}$. We now show how to reconstruct the information we need for constructing posterior predictive tests using the information we do have access to.  
 
 \codefont{BORG} samples the posterior for initial density and bias parameters, which are combined to give $\bar{\lambda}_i^{\mathrm{c}}=\lambda_i^{\mathrm{c}}/A_{\alpha(i)}^{\mathrm{c}}$. If we have a healpix region containing voxels $i=1\ldots I$, with amplitude $A_{\alpha}^{\mathrm{c}}$ in catalogue $\mathrm{c}$, we can compute the joint posterior distribution for $\bar{\lambda}_{i}^{\mathrm{c}}$ and $A_{\alpha}^{\mathrm{c}}$ conditioned on  $N^{\mathrm{c}}_{1\ldots I}$:
	\begin{align}
		P(\bar{\lambda}_{1\ldots I}^{\mathrm{c}}&,A_{\alpha}^{\mathrm{c}}|N_{1\ldots I}^{\mathrm{c}}) \nonumber \\ =&P(\bar{\lambda}_{1\ldots I}^{\mathrm{c}}|N_{1\ldots I}^{\mathrm{c}})P(A_{\alpha}^{\mathrm{c}}|\bar{\lambda}_{1\ldots I}^{\mathrm{c}},N_{1\ldots I}^{\mathrm{c}}) \nonumber \\
		=& P(\bar{\lambda}_{1\ldots I}^{\mathrm{c}}|N_{1\ldots I}^{\mathrm{c}}) \frac{P(A_{\alpha}^{\mathrm{c}}|\bar{\lambda}_{1\ldots I}^{\mathrm{c}})P(N_{1\ldots I}^{\mathrm{c}}|A_{\alpha}^{\mathrm{c}},\bar{\lambda}_{1\ldots I}^{\mathrm{c}})}{P(N_{1\ldots I}^{\mathrm{c}}|\bar{\lambda}_{1\ldots I})} \nonumber \\
		=&P(\bar{\lambda}_{1\ldots I}^{\mathrm{c}}|N_{1\ldots I}^{\mathrm{c}}) \nonumber \\
  &\times \frac{P(A_{\alpha}^{\mathrm{c}})\prod_{i=1}^{I}(\bar{\lambda}_i^{\mathrm{c}}A_{\alpha}^{\mathrm{c}})^{N_i^{\mathrm{c}}}e^{-\bar{\lambda}_{i}^{\mathrm{c}}A_{\alpha}^{\mathrm{c}} }/N_i^{\mathrm{c}}!}{\int \mathrm{d}A_{\alpha}^{'\mathrm{c}}P(A_{\alpha}^{'\mathrm{c}})\prod_{i=1}^{I}(\bar{\lambda}_{i}^{\mathrm{c}}A_{\alpha}^{'\mathrm{c}})^{N_i^{\mathrm{c}}}e^{-\bar{\lambda}_{i}^{\mathrm{c}}A_{\alpha}^{'\mathrm{c}}}/{N_i^{\mathrm{c}}}!}
	\end{align}

Here we have used Bayes theorem in the second step, and assumed conditional independence between $A_{\alpha}^{\mathrm{c}}$ and $\bar{\lambda}_{1\ldots I}^{\mathrm{c}}$ in the third. We have also made use of Eq.~(\ref{eq:Poisson_like}). Using the same Jeffreys prior as before, $P(A_{\alpha}^{\mathrm{c}}) = \kappa_{\alpha}/A_{\alpha}^{\mathrm{c}}$, and defining $N_{\mathrm{tot},\alpha}^{\mathrm{c}} = \sum_{i=1}^IN_{i}^{\mathrm{c}}$ and $\bar{\lambda}_{\mathrm{tot},\alpha}^{\mathrm{c}} = \sum_{i=1}^{I}\bar{\lambda}_{i}^{\mathrm{c}}$ we find
	\begin{align}
		P(\bar{\lambda}_{1\ldots I}^{\mathrm{c}}&,A_{\alpha}^{\mathrm{c}}|N_{1\ldots I}^{\mathrm{c}}) \nonumber \\
		=& P(\bar{\lambda}_{1\ldots I}^{\mathrm{c}}|N_{1\ldots I}^{\mathrm{c}}) \frac{1}{A_{\alpha}^{\mathrm{c}}} \frac{(A_{\alpha}^{\mathrm{c}}\bar{\lambda}_{\mathrm{tot},\alpha}^{\mathrm{c}})^{N_{\mathrm{tot},\alpha}^{\mathrm{c}}}e^{-A_{\alpha}^{\mathrm{c}}\bar{\lambda}_{\mathrm{tot},\alpha}^{\mathrm{c}}}}{\Gamma(N_{\mathrm{tot},\alpha}^{\mathrm{c}})}\, .
	\end{align}
	The \codefont{BORG} Markov chain provides samples from the posterior distribution $P(\bar{\lambda}_{1\ldots I}^{\mathrm{c}}|N_{1\ldots I}^{\mathrm{c}})$. By marginalising over $\bar{\lambda}_{1\ldots I}^{\mathrm{c}}$ using Monte-Carlo integration, we can estimate the posterior distribution for $A_{\alpha}^{\mathrm{c}}$:
	\begin{align}
		P(A_{\alpha}^{\mathrm{c}}&|N_{1\ldots I}^{\mathrm{c}}) \nonumber \\
		\approx & \frac{(A_{\alpha}^{\mathrm{c}})^{{N^{\mathrm{c}}_{\mathrm{tot},\alpha}-1}}}{{\left(N_{\mathrm{tot},\alpha}^{\mathrm{c}}-1\right)!}} \frac{1}{S}\sum_{s=1}^{S}	 (\bar{\lambda}_{\mathrm{tot},\alpha,s}^{\mathrm{c}})^{N_{\mathrm{tot},\alpha}^{\mathrm{c}}}e^{-A_{\alpha}^{\mathrm{c}}\bar{\lambda}_{\mathrm{tot},\alpha,s}^{\mathrm{c}}} \, .
	\end{align}
	Here, we have taken $S$ samples from the posterior, and $\bar{\lambda}_{\mathrm{tot},\alpha,s}^{\mathrm{c}}$ is calculated from the MCMC sample $s$. We therefore obtain the expectation value of $A_{\alpha}^{\mathrm{c}}$ as 
	\begin{align}
		E(A_{\alpha}^{\mathrm{c}}|N_{1\ldots I}^{\mathrm{c}}) =& \int_{0}^{\infty}\mathrm{d}A_{\alpha}^{\mathrm{c}} A_{\alpha}^{\mathrm{c}} P(A_{\alpha}^{\mathrm{c}}|N_{1\ldots I}^{\mathrm{c}}) \nonumber \\
		\approx& \frac{N_{\mathrm{tot},\alpha}^{\mathrm{c}}}{S}\sum_{s=1}^{S} \frac{1}{\bar{\lambda}_{\mathrm{tot},\alpha,s}^{\mathrm{c}}}\, ,
	\end{align}
	yielding Eq.~(\ref{eq:Aalpha}). An analogous computation gives the expectation value for the mean number of galaxies in pixel $i$, $A_{\alpha}^{\mathrm{c}}\bar{\lambda}_{i}^{\mathrm{c}}$, yielding Eq.~(\ref{eq:Elambdai}):
	\begin{equation}
	E(A_{\alpha}^{\mathrm{c}}\bar{\lambda}_{i}|N_{1\ldots I}^{\mathrm{c}}) \approx \frac{N_{\mathrm{tot},\alpha}^{\mathrm{c}}}{S}\sum_{s=1}^S\frac{\bar{\lambda}_{i,s}^{\mathrm{c}}}{\bar{\lambda}_{\mathrm{tot},\alpha,s}^{\mathrm{c}}}\, .
	\end{equation}

	\section{Posterior Predictive Tests for Other Clusters}
	\label{app:ppts_other}
	
\begin{figure*}
		\centering
		\includegraphics[width=\textwidth]{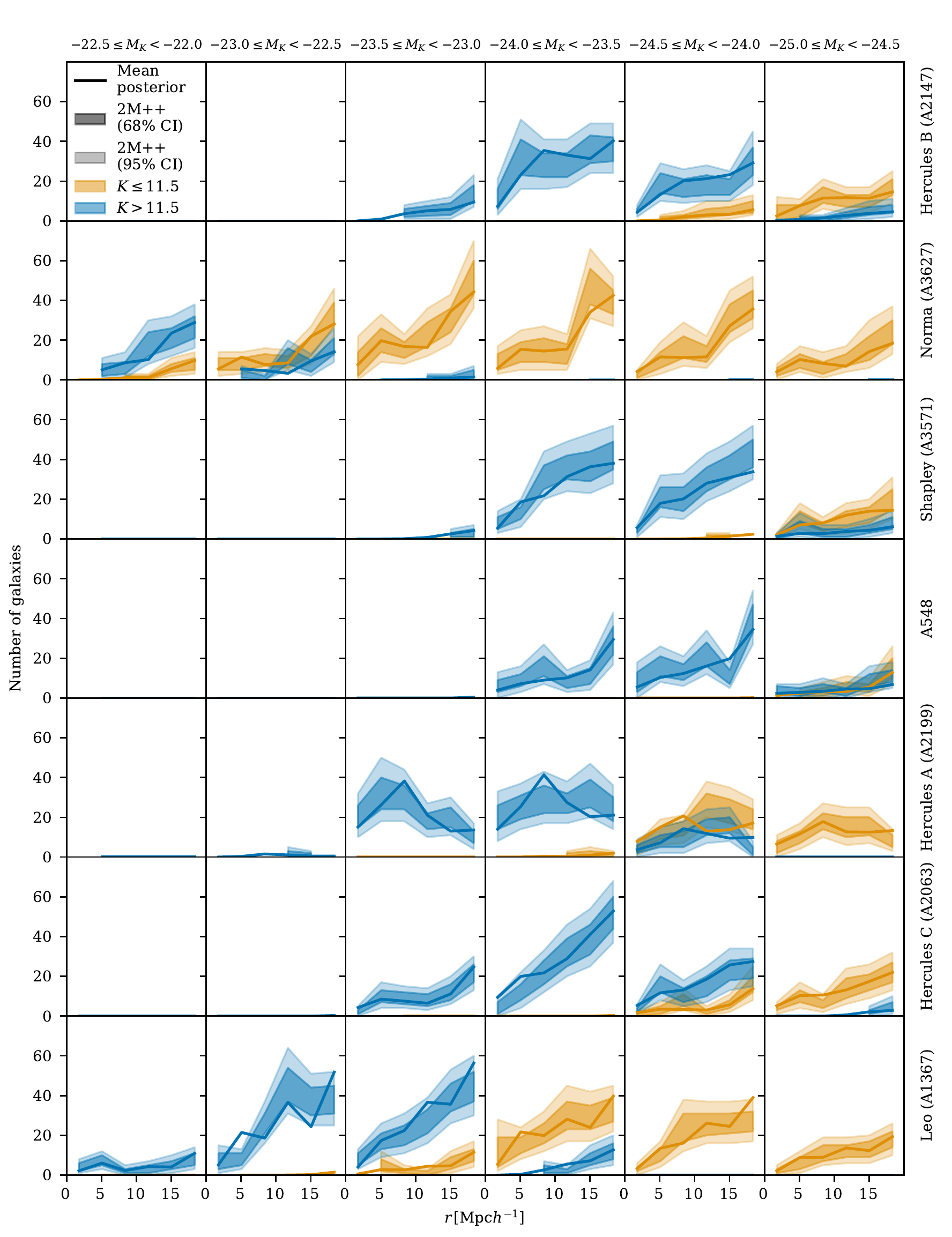}
		\caption{\label{fig:ppts_other} Posterior predictive tests for the other seven clusters considered in Fig.~\ref{fig:cluster_masses}, using the same conventions as Fig.~\ref{fig:ppt}}
	\end{figure*}

	We show in Fig.~\ref{fig:ppts_other} the posterior predictive tests for the remaining seven clusters considered in this work. For all these clusters, the posterior predictive tests pass in all but a handful of radial bins.
	
	\interlinepenalty=10000

	\bibliographystyle{mnras}
	\bibliography{borg_antihalos_paper}
\label{lastpage}
	
\end{document}